\documentclass{article}

\usepackage[preprint]{cpal_2026}

\usepackage{graphicx}   % for \resizebox
\usepackage{booktabs}   % for \toprule, \midrule, \bottomrule
\usepackage{wrapfig} 
\usepackage[table]{xcolor} % for \rowcolor 
\usepackage{arydshln}
\usepackage{pifont}
\usepackage{tcolorbox}
\usepackage{url}
\usepackage{amsmath,amssymb,amsfonts}
\usepackage{cleveref}

\newcommand{\myparagraph}[1]{\vspace{0pt}\noindent{\bf #1}}

\title{SonoEdit: Null-Space Constrained Knowledge Editing for Pronunciation Correction in LLM-Based TTS}

\author{%
  Ayush Pratap Singh\textsuperscript{1}, Harshit Singh\textsuperscript{2}, Nityanand Mathur\textsuperscript{3}, Akshat Mandloi\textsuperscript{3}, Sudarshan Kamath\textsuperscript{3} \\
  \textsuperscript{1}TU Darmstadt, \textsuperscript{2}UMD, \textsuperscript{3}Smallest AI\\
  \texttt{nityanand@smallest.ai}
}

\begin{document}

\maketitle

\begin{abstract}
Neural text-to-speech systems systematically mispronounce low-resource proper nouns, particularly non-English names, brands, and geographic locations due to their underrepresentation in predominantly English training corpora. Existing solutions require expensive multilingual data collection or manual phonetic annotation, limiting TTS deployment in diverse linguistic contexts. We introduce \textbf{SonoEdit}, a model editing technique that surgically corrects pronunciation errors in pre-trained TTS models without retraining. Correcting such errors traditionally requires costly supervised finetuning or manual phoneme injection. In this work, we present a parsimonious alternative using Null-Space Pronunciation Editing, a single-shot parameter update that modifies the pronunciation of specific words while provably preserving the rest of the model’s behavior. We first adapt Acoustic Causal Tracing to identify the specific Transformer layers governing text-to-pronunciation mapping. We then employ Null-Space Constrained Editing to compute a closed-form weight update that rectifies the target pronunciation while remaining mathematically orthogonal to the manifold of general speech, constructing a constrained update that drives the model’s acoustic output toward a desired pronunciation exemplar while ensuring zero first-order change on a preserved corpus.
\end{abstract}

\section{Introduction}

Modern neural text-to-speech systems have achieved remarkable naturalness and expressiveness, yet they consistently struggle with a fundamental challenge: correctly pronouncing words outside their training distribution. Proper nouns (brand names, personal names, geographic locations), technical terminology, and neologisms are frequently mispronounced, limiting the deployment of TTS in domains requiring accurate pronunciation like customer service, navigation systems, educational content, and accessibility applications. Phoneme dictionaries~\cite{kominek2004cmu} require manual annotation and don't generalize to morphological variations. 
Traditional TTS systems like Tacotron 2~\cite{betker2022tortoise} and FastSpeech 2~\cite{ren2021fastspeech2} handle pronunciations 
through explicit phoneme inputs. Recent codec-based systems like VALL-E~\cite{wang2023neural} and VoiceBox~\cite{le2023voicebox} learn implicit mappings but struggle with rare words.
Global fine-tuning is prohibitively expensive and prone to catastrophic forgetting, where correcting a single pronunciation can degrade prosody, speaker identity, or other acoustic properties. LoRA and similar PEFT methods reduce computational cost but still require task-specific training and lack explicit guarantees about preserving general model behavior. Moreover, prompt engineering for LLM-based TTS is unreliable and context-dependent. All these leads to a very important question in terms of text-speech alignment:

\begin{tcolorbox}[colback=gray!20, colframe=gray!40, boxrule=0.5pt, arc=2mm, left=3mm, right=3mm, top=1.5mm, bottom=1.5mm]
\textit{Where does pronunciation live in a large language model(LLM) based text-to-speech(TTS) system, and how can we surgically modify it for specific texts while preserving all other model behavior?}
\end{tcolorbox}

SonoEdit addresses this very problem through a novel combination of three key ideas:

\textbf{1. Precise Localization via Acoustic Causal Tracing:} By adapting causal tracing techniques from the mechanistic interpretability literature, we identify the specific Transformer layers (often a narrow range such as layers $i$ to $j$) that are responsible for handling grapheme-to-phoneme mappings. This localization is more precise than layer-wise or module-level interventions, allowing us to focus corrections exactly where they are needed.\\
\textbf{2. Orthogonal Corrections via Null-Space Constrained Editing:} Rather than applying arbitrary weight perturbations or training new parameters, we compute a closed-form weight update that rectifies the target pronunciation while remaining mathematically orthogonal to the manifold of general speech. Specifically, we identify the null-space of the projection operator corresponding to general acoustic features, and constrain our weight update to lie entirely within this null-space. This ensures that the correction cannot affect any aspect of model behavior other than the target pronunciation.\\
\textbf{3. Parameter Parsimony and One-Shot Correction:} Unlike PEFT methods, SonoEdit does not introduce additional trainable parameters. The weight update is computed in closed-form via simple matrix operations, making it computationally efficient and applicable in a one-shot manner without any training loop. This is particularly valuable for deployed TTS systems where re-training is infeasible.\\
\textbf{To the best of our knowledge, SonoEdit is the first work to explore the application of null-space knowledge editing to neural TTS systems.} By synthesizing precise localization, orthogonal constraints, and closed-form solutions, SonoEdit overcomes the current trade-off between computational efficiency, correctness preservation, and parameter efficiency. It enables surgical, parsimonious modifications to LLM-based TTS models that are guaranteed not to introduce catastrophic forgetting while requiring minimal computational overhead.
\section{Related Works}

\subsection{LLM-based Text-to-Speech Modelling}

Recent advances in neural text-to-speech (TTS) have demonstrated that Large Language Models (LLMs) can be effectively repurposed for high-quality speech synthesis. 
AudioLM~\cite{borsos2023audiolm} pioneered language modeling for 
audio generation, while VALL-E~\cite{wang2023neural} demonstrated zero-shot voice 
cloning through in-context learning. SoundStorm~\cite{borsos2023soundstorm} improved 
efficiency through parallel generation, and NaturalSpeech 2 ~\cite{shen2023naturalspeech2} 
achieved human-parity through latent diffusion. These systems leverage neural 
codecs like EnCodec~\cite{defossez2022high} and SoundStream~\cite{zeghidour2021soundstream} to discretize audio into tokens.Models such as Orpheus~\cite{orpheus-tts2025} and Sesame~\cite{sesame2025uncannyvalley} leverage the semantic and linguistic understanding learned during pretraining to generate expressive, natural-sounding speech with superior prosody compared to traditional concatenative or unit-selection methods.

These LLM-based TTS approaches operate by conditioning the speech decoder on embeddings derived from text or other linguistic representations. The key advantage is that LLMs capture long-range linguistic dependencies and tonal variations, enabling the generation of speech with nuanced prosody, stress patterns, and intonation contours. However, despite their remarkable performance on in-distribution data, these models exhibit a critical failure mode: the hallucination of incorrect pronunciations for rare or out-of-distribution proper nouns that were rarely encountered during training. This phenomenon is particularly problematic in applications such as navigation systems, news reading, or multilingual voice assistants, where accurate pronunciation of named entities is essential for user comprehension.
\definecolor{mylightgreen}{HTML}{82B366}
\definecolor{mylightorange}{HTML}{D79B00}
\definecolor{mypurple}{HTML}{9673A6}
\begin{figure}
    \centering
    \includegraphics[width=\linewidth]{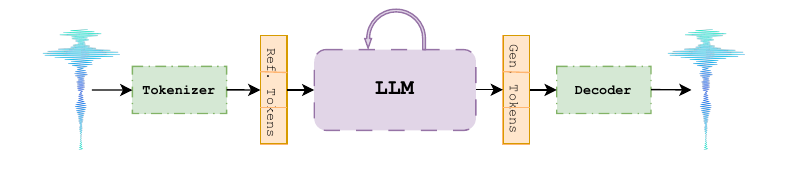}
    \caption{A schematic diagram illustrating an LLM-based speech generation pipeline. A \textcolor{cyan}{Raw Waveform} is first processed by a \textcolor{mylightgreen}{Tokenizer} to produce a sequence of \textcolor{mylightorange}{Discrete Tokens}. These tokens are then input into a \textcolor{mypurple}{Large Language Model (LLM)}, which generates a sequence of \textcolor{mylightorange}{New Tokens}. Finally, a \textcolor{mypurple}{Decoder} converts the new tokens into a \textcolor{cyan}{Speech Waveform}.}
    \label{fig:llm-tts}
\end{figure}

\subsection{Locating and Editing Word-Pronunciation Associations in TTS}

Understanding and modifying the internal representations of neural TTS models is a critical research direction. Prior work has explored localization through probing classifiers~\cite{belinkov2019analysis}, causal tracing~\cite{meng2022locating}, and attribution patching~\cite{geiger2023causal}. Layer-wise analysis~\cite{tenney2019bert} in language models revealed that different layers encode different linguistic phenomena.

In the context of TTS, recent work has focused on identifying which components of the model are responsible for handling token-to-pronunciation features. Some studies have proposed techniques to directly intervene in specific layers to correct pronunciation errors. However, such approaches often employ crude modification strategies, such as simple weight perturbations or layer-wise interventions, which lack the precision necessary to modify target pronunciations without degrading the model's general acoustic capabilities or speaker identity preservation.

Our work builds upon these insights by employing a more principled approach: we use Acoustic Causal Tracing to precisely identify the Transformer layers responsible for handling G2P mappings, and then apply null-space constrained editing to ensure that corrections are orthogonal to the manifold of general speech characteristics.

\subsection{Knowledge Editing}

Knowledge editing is an emerging paradigm for modifying the factual knowledge encoded in neural networks without catastrophic forgetting. Recent methods such as ROME (Rank-One Model Editing)~\cite{rome}, MEMIT~\cite{meng2022mass}, MEND~\cite{mitchell2021fast}, SERAC~\cite{mitchell2021fast}, and AlphaEdit~\cite{fang2024alphaedit} have shown promise in correcting facts in Large Language Models while preserving the model's general capabilities.

The core insight of these methods is that knowledge in neural networks is often localized to specific layers or subspaces. By operating in these subspaces particularly the null-space of the forward projection, interventions can be made orthogonal to existing model behavior, thereby minimizing side effects. ROME, for instance, uses rank-one updates to correct factual associations in LLMs by computing weight changes that preserve the rest of the model's output distribution.

While these methods were originally designed for factual knowledge correction in language models, their underlying principles are highly relevant to TTS pronunciation correction. Just as a language model encodes facts in localized, structured representations, a TTS model encodes pronunciation mappings in specific layers. By adapting causal tracing and null-space constrained editing techniques from the knowledge editing literature, we can apply principled, parsimonious interventions to correct pronunciation errors while maintaining acoustic fidelity and speaker identity. This cross-domain adaptation is a key contribution of our work.

\subsection{Parameter Efficient Fine-tuning}

Full model fine-tuning is computationally expensive and prone to catastrophic forgetting, especially when correcting errors in specific, localized aspects of model behavior. To address this, parameter-efficient fine-tuning (PEFT) methods such as LoRA~\cite{hu2022lora}, adapters~\cite{houlsby2019parameter}, QLoRA~\cite{dettmers2023qlora}, and Prefix Tuning~\cite{li2021prefix} have gained widespread adoption.

LoRA, one of the most popular PEFT methods, assumes that weight updates lie in a low-rank subspace. Rather than updating all parameters, LoRA introduces small trainable matrices $A \in \mathbb{R}^{r \times d_{in}}$ and $B \in \mathbb{R}^{d_{out} \times r}$ where $r \ll d_{in}, d_{out}$, and updates are computed as $\Delta W = B A$. This reduces the number of trainable parameters while often preserving model performance.

However, LoRA and similar PEFT methods have significant limitations in the context of pronunciation correction. First, they require training data specific to the target pronunciation errors, which may not always be readily available. Second, even with careful hyperparameter tuning, these methods can introduce subtle artifacts or degradations in acoustic quality when applied to locally-constrained corrections. Third, they do not explicitly enforce orthogonality to the manifold of general speech, so corrections can still indirectly interfere with other aspects of model behavior.

In contrast, our approach (SonoEdit) operates directly in the null-space of general speech characteristics, ensuring that pronunciation corrections are mathematically guaranteed to be orthogonal to the model's core acoustic capabilities. By combining causal tracing with null-space constrained editing, we achieve a more principled and parsimonious solution that requires neither expensive retraining nor introduction of additional parameters.

\section{Methodology}

\subsection{Preliminaries}

\paragraph{LLMs for Speech Planning}
We consider LLM-based text-to-speech (TTS) in which a Large Language Model (LLM) serves as an intermediate planner between text and acoustics. In architectures like Orpheus, the LLM generates a flattened sequence of discrete acoustic tokens $\mathbf{t} = [c_{1}^{(1)}, \dots, c_{1}^{(7)}, c_{2}^{(1)}, \dots, c_{N}^{(7)}]$, where each acoustic frame is represented by 7 hierarchical tokens derived from a neural audio codec (SNAC \cite{2024arXiv241014411S}). Here, $c_{i}^{(1)}$ represents the coarse semantic/acoustic token for frame $i$, carrying the primary phonological content, while $c_{i}^{(2\dots7)}$ represent fine acoustic details such as pitch contours, voicing characteristics, and spectral fine structure. The generative process can be formalized as a sequential prediction task where the LLM models the conditional distribution:
\begin{equation}
p(\mathbf{t} \mid \mathbf{x}) = \prod_{i=1}^{N} \prod_{j=1}^{7} p(c_i^{(j)} \mid c_{<i}, c_i^{(<j)}, \mathbf{x})
\end{equation}
where $\mathbf{x}$ denotes the input text, $c_{<i}$ represents all tokens from previous frames, and $c_i^{(<j)}$ denotes tokens within the current frame preceding level $j$. Since phonological planning decisions are primarily encoded in the coarse tokens, we restrict our causal analysis and editing to the subspace of coarse tokens $\{c_{i}^{(1)}\}_{i=1}^N$. Crucially, pronunciation errors in these systems do not originate from the decoder, but from incorrect internal associations formed within the LLM's coarse planning. Therefore, correcting pronunciation requires modifying the LLM's internal representations rather than retraining or adapting the acoustic components of the system.

% \myparagraph{LLMs for Speech Planning: }
% We consider LLM-based text-to-speech (TTS) in which a Large Language Model (LLM) serves as an intermediate planner between text and acoustics. In architectures like Orpheus, the LLM generates a flattened sequence of discrete acoustic tokens $\mathbf{t} = [c_{1}^{(1)}, \dots, c_{1}^{(7)}, c_{2}^{(1)}, \dots, c_{N}^{(7)}]$, where each acoustic frame is represented by 7 hierarchical tokens derived from a neural audio codec (SNAC \cite{2024arXiv241014411S}). Here, $c_{i}^{(1)}$ represents the coarse semantic/acoustic token for frame $i$, carrying the primary phonological content, while $c_{i}^{(2\dots7)}$ represent fine acoustic details.

% Since phonological planning decisions are primarily encoded in the coarse tokens, we restrict our causal analysis and editing to the subspace of coarse tokens $\{c_{i}^{(1)}\}_{i=1}^N$. Crucially, pronunciation errors in these systems do not originate from the decoder, but from incorrect internal associations formed within the LLM's coarse planning. Therefore, correcting pronunciation requires modifying the LLM’s internal representations rather than retraining or adapting the acoustic components of the system.

\paragraph{Knowledge Editing in LLMs: }
Knowledge editing frames model correction as a targeted modification of internal parameters that alters a specific input–output association while preserving all other behaviors. Given a model with parameters $W$, the goal is to compute a minimal update $\Delta W$ such that the model produces a desired output for a particular input, without affecting its responses elsewhere.
\begin{equation}
\Delta W^* = \arg\min_{\Delta W} \left\| (W + \Delta W) K_1 - V_1 \right\|_F^2 + \lambda \left\| (W + \Delta W) K_0 - V_0 \right\|_F^2
\end{equation}
where $K_1, V_1$ represent keys and values for the target pronunciation, and $K_0, V_0$ represent keys and values for preserved knowledge. The parameter $\lambda$ balances the trade-off between correcting the target and preserving existing knowledge.

Recent work has shown that many associations in LLMs are localized to specific layers and subspaces, enabling closed-form, one-shot edits that avoid catastrophic forgetting. In this work, we treat incorrect pronunciations as faulty associative memories embedded in the LLM’s hidden representations and apply knowledge editing techniques to correct them in a principled and parsimonious manner.

\subsection{Acoustic Causal Tracing}
Pronunciation-related information is not uniformly distributed across an LLM. Instead, it is concentrated in a narrow subset of Transformer layers responsible for mapping orthographic forms to latent phonological representations.

To identify these layers, we adapt causal tracing techniques from mechanistic interpretability to the flattened SNAC architecture. We perform interventional analysis by corrupting the input representations via noise injection and selectively restoring individual layer activations from a clean forward pass. Specifically, we define the \emph{Acoustic Causal Impact} of a layer $\ell$ as the recovery in probability of the correct coarse token $c^*$ when the layer's activations are restored from a clean run:
\begin{equation}
\text{Impact}(\ell) = \mathbb{P}[c^* \mid \text{do}(h^{(\ell)} = h_{\text{clean}}^{(\ell)})] - \mathbb{P}[c^* \mid \text{corrupted}]
\end{equation}
where $c^*$ is the first (coarse) token of the SNAC frame corresponding to the mispronounced phoneme, $h^{(\ell)}$ denotes the hidden state at layer $\ell$, and the $\text{do}(\cdot)$ operator represents a causal intervention. For computational efficiency, we estimate these probabilities using the model's logits:
\begin{equation}
\text{Impact}(\ell) \approx \frac{\exp(z_{\text{restored}}^{(\ell)}[c^*])}{\sum_{c'} \exp(z_{\text{restored}}^{(\ell)}[c'])} - \frac{\exp(z_{\text{corrupted}}[c^*])}{\sum_{c'} \exp(z_{\text{corrupted}}[c'])}
\end{equation}
where $z^{(\ell)}$ represents the logit vector after layer $\ell$. Layers with high impact scores are identified as causally responsible for word-pronunciation mappings. Empirically, we find that pronunciation-sensitive representations are localized to a contiguous block of mid-to-late Transformer layers (typically layers $L/2$ to $3L/4$ for a model with $L$ layers). This precise localization allows us to restrict subsequent edits to a small set of parameters, avoiding unnecessary interference with semantic planning or acoustic rendering.

% \subsection{Acoustic Causal Tracing}
% Pronunciation-related information is not uniformly distributed across an LLM. Instead, it is concentrated in a narrow subset of Transformer layers responsible for mapping orthographic forms to latent phonological representations.

% To identify these layers, we adapt causal tracing techniques from mechanistic interpretability to the flattened SNAC architecture. Specifically, we define the \emph{Acoustic Causal Impact} of a layer $\ell$ as the recovery in probability of the correct coarse token $c^*$ when the layer's activations are restored from a clean run:
% \begin{equation}
% \text{Impact}(\ell) = \mathbb{P}[c^* \mid \text{do}(h^{(\ell)} = h_{\text{clean}})] - \mathbb{P}[c^* \mid \text{corrupted}]
% \end{equation}
% where $c^*$ is the first (coarse) token of the SNAC frame corresponding to the mispronounced phoneme. Layers with high impact scores are identified as causally responsible for word-pronunciation mappings. Empirically, we find that pronunciation-sensitive representations are localized to a contiguous block of mid-to-late Transformer layers. This precise localization allows us to restrict subsequent edits to a small set of parameters, avoiding unnecessary interference with semantic planning or acoustic rendering.

\subsection{Null Space: Importance and Utilization}

A central challenge in pronunciation correction is ensuring that edits do not degrade general speech characteristics such as prosody, speaker identity, or fluency. To address this, we explicitly characterize the subspace corresponding to general speech behavior and constrain our edits to be orthogonal to it. Let $K_0 \in \mathbb{R}^{d \times N}$ be a matrix of hidden states (keys) collected from the prediction steps of coarse tokens across a diverse speech dataset (e.g., LibriTTS\cite{zen2019libritts}). We compute the uncentered covariance matrix $\Sigma = K_0 K_0^T$ to capture the manifold of general speech representations.
\begin{wrapfigure}{lh}{0.5\linewidth}
    \centering
    \vspace{-10pt}
    \includegraphics[width=\linewidth]{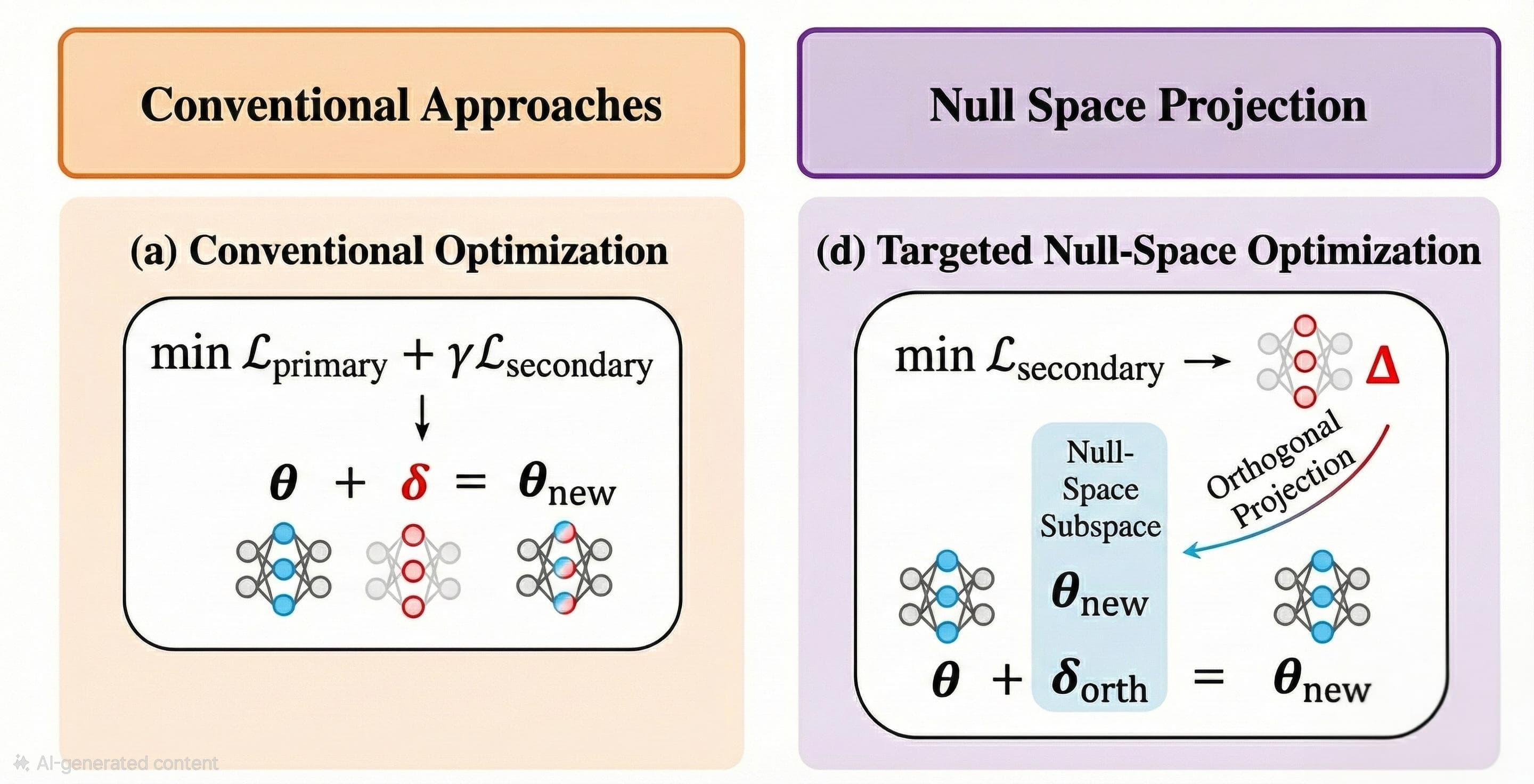}
    \caption{An illustration of null-space constrained editing. The pronunciation correction is applied in directions orthogonal to the subspace representing general speech characteristics, ensuring invariance of non-target behavior.}
    \label{fig:placeholder}
    \vspace{-10pt}
\end{wrapfigure}
The null space of this manifold consists of directions that do not affect general speech behavior. We compute the orthonormal basis of this null space using Singular Value Decomposition (SVD) on the covariance matrix $\Sigma$. The projection matrix $P$ onto the null space is given by:
\begin{equation}
P = I - U U^T
\end{equation}
where $U$ contains the eigenvectors of $\Sigma$ corresponding to the dominant eigenvalues (the "speech manifold"). Any weight update $\Delta W$ satisfying $\Delta W = \Delta W P$ is guaranteed to have minimal impact on general speech capabilities, as $\Delta W k \approx 0$ for any $k$ in the effective range of $K_0$.

% \subsection{Null-Space Constrained Knowledge Editing}
% Once pronunciation-relevant layers and the null space of general speech have been identified, we compute the final weight update via a constrained optimization objective. Given a target subject key $k_*$ (the hidden state at the error location) and a target value $v_*$ (the vector that induces the correct coarse token), we solve for the weight update $\Delta W$ that minimizes the error on the target while remaining orthogonal to preserved knowledge.

% Using the AlphaEdit\cite{fang2024alphaedit} estimator, this optimization admits a closed-form solution:
% \begin{equation}
% \Delta W = \frac{v_* - W k_*}{k_*^T P k_*} (P k_*)^T
% \end{equation}
% This update ensures that $(W + \Delta W)k_* = v_*$ while strictly adhering to the null-space constraint $\Delta W K_0 \approx 0$. The resulting update is applied only to the localized layers identified via acoustic causal tracing.

% As a result, SonoEdit enables one-shot pronunciation correction: a single computation permanently fixes the target pronunciation while leaving all other pronunciations, speakers, and prosodic patterns unchanged. This makes the approach particularly well-suited for deployed TTS systems where retraining or parameter expansion is impractical.
\subsection{Null-Space Constrained Knowledge Editing}

Once pronunciation-relevant layers and the null space of general speech have been identified, we compute the final weight update via a constrained optimization objective. Given a target subject key $k_* \in \mathbb{R}^{d}$ (the hidden state at the error location) and a target value $v_* \in \mathbb{R}^{d}$ (the vector that induces the correct coarse token), we solve for the weight update $\Delta W$ that minimizes the error on the target while remaining orthogonal to preserved knowledge:
\begin{equation}
\min_{\Delta W} \left\| (W + \Delta W)k_* - v_* \right\|_2^2 \quad \text{subject to} \quad \Delta W K_0 = 0
\end{equation}
where the constraint ensures that the update produces zero perturbation on the general speech distribution represented by $K_0$. Using the AlphaEdit\cite{fang2024alphaedit} estimator, this optimization admits a closed-form solution:
\begin{equation}
\Delta W = \frac{v_* - W k_*}{k_*^T P k_*} (P k_*)^T
\end{equation}
where the numerator $(v_* - W k_*)$ represents the residual error and the denominator $k_*^T P k_* = \|P k_*\|_2^2$ normalizes by the null-space component magnitude. This update ensures that $(W + \Delta W)k_* = v_*$ while strictly adhering to the null-space constraint $\Delta W K_0 \approx 0$. To verify constraint satisfaction, we note that:
\begin{equation}
\Delta W K_0 = \frac{v_* - W k_*}{k_*^T P k_*} (P k_*)^T K_0 = \frac{v_* - W k_*}{k_*^T P k_*} k_*^T P K_0 \approx 0
\end{equation}
since $P K_0 = (I - U_k U_k^T) K_0 \approx 0$ by construction—the columns of $K_0$ lie predominantly in the span of $U_k$. The resulting update is applied only to the localized layers identified via acoustic causal tracing, specifically to the value projection matrices $W_V^{(\ell)}$ or feed-forward layers $W_{\text{FF}}^{(\ell)}$ at layer $\ell \in \mathcal{L}_{\text{edit}}$. As a result, SonoEdit enables one-shot pronunciation correction: a single computation permanently fixes the target pronunciation while leaving all other pronunciations, speakers, and prosodic patterns unchanged. The entire process requires only two forward passes to extract keys and values, a precomputed projection matrix $P$ (obtained via one-time SVD), and $O(d^2)$ operations for the rank-1 weight update. This makes the approach particularly well-suited for deployed TTS systems where retraining or parameter expansion is impractical, enabling rapid post-deployment corrections without model degradation.
\section{Experiments}
\label{sec:experiments}
Our experiments aim to address three major research questions fundamental to the SonoEdit approach: \ding{182} can null-space constrained edits reliably and precisely correct pronunciation for low-resource proper nouns without requiring acoustic decoder retraining? \ding{183} do these edits preserve the global behavior of Orpheus-TTS with respect to prosody, semantics, and fluency? And \ding{184} how small and parsimonious can an edit be while remaining effective in correcting target pronunciations? To investigate these questions comprehensively, we conduct all experiments in a one-shot editing regime, without any retraining of the acoustic decoder.

% All experiments are conducted in a \emph{one-shot editing} regime, without retraining the acoustic decoder.

\subsection{Implementation Details}

\myparagraph{Evaluation Models: }
We evaluate SonoEdit on Orpheus-TTS~\cite{orpheus-tts2025}, which leverages a LLaMA-3B~\cite{grattafiori2024llama} backbone for semantic understanding and speech generation, and Sesame-TTS. To implement SonoEdit, we employ Acoustic Causal Tracing to identify the specific Transformer layers responsible for grapheme-to-phoneme mappings. This analysis reveals that layers 16--22 are most causally responsible for text-pronunciation behavior, providing precise localization compared to cruder layer-wise interventions. Once these layers are identified, SonoEdit computes the null-space using a subset of LibriTTS~\cite{zen2019libritts}. Critically, all edits are performed once, using a closed-form matrix update without iterative optimization, enabling one-shot application in deployed systems.

\myparagraph{Evaluation Data: } 
To rigorously test pronunciation correction across diverse linguistic contexts, we construct HardNoun-300, a carefully curated multilingual dataset of 300 proper nouns (50 per language across 6 languages: English, Spanish, French, German, Japanese, and Hindi) that consistently induce hallucinations in pretrained LLM-based TTS systems. Our selection criteria prioritize words that are (1) systematically mispronounced by the baseline Orpheus model with error rates exceeding 80\%, (2) phonetically unambiguous with clear ground-truth pronunciations verified by native speakers, and (3) representative of real-world deployment challenges in multilingual applications. 
Each language-specific subset spans four categories: personal names, geographic locations, international brand names, and culturally significant terms. For each of the 300 target nouns, we construct 10 contextually diverse sentences varying in syntactic position, semantic context, and prosodic structure yielding 3,000 total test utterances. Reference pronunciations are obtained through two complementary methods: for well-documented words, we use phoneme-injected synthesis by manually specifying IPA transcriptions and generating audio through a phoneme-aware TTS system. For culturally sensitive terms, we collect recordings from native speakers to ensure authenticity and natural coarticulation patterns.

\myparagraph{Evaluation Metrics: }
For evaluation, we employ both objective and subjective metrics to comprehensively assess the effectiveness of SonoEdit. To measure target pronunciation accuracy, we compute two complementary metrics. Target-WER evaluates the Word Error Rate on the target noun pronunciation, while Phoneme Error Rate (PER) provides fine-grained assessment of phoneme-level accuracy through forced alignment. To assess global model preservation, Global-WER measures the Word Error Rate on the preservation set, ensuring that edits do not degrade model performance on unrelated utterances. Finally, we use Speaker Similarity (SIM), computed as the cosine similarity of WavLM~\cite{chen2022wavlm} speaker embeddings, to verify that speaker identity and acoustic characteristics are preserved through the editing process. Additionally, we evaluate MOS scores using UTMOS~\cite{saeki2022utmos} which captures perceptual acoustic quality beyond objective metrics.

\subsection{Layer Sensitivity Analysis} \label{ssec:layer}
\begin{figure}[h]
    \centering
    \includegraphics[width=\linewidth]{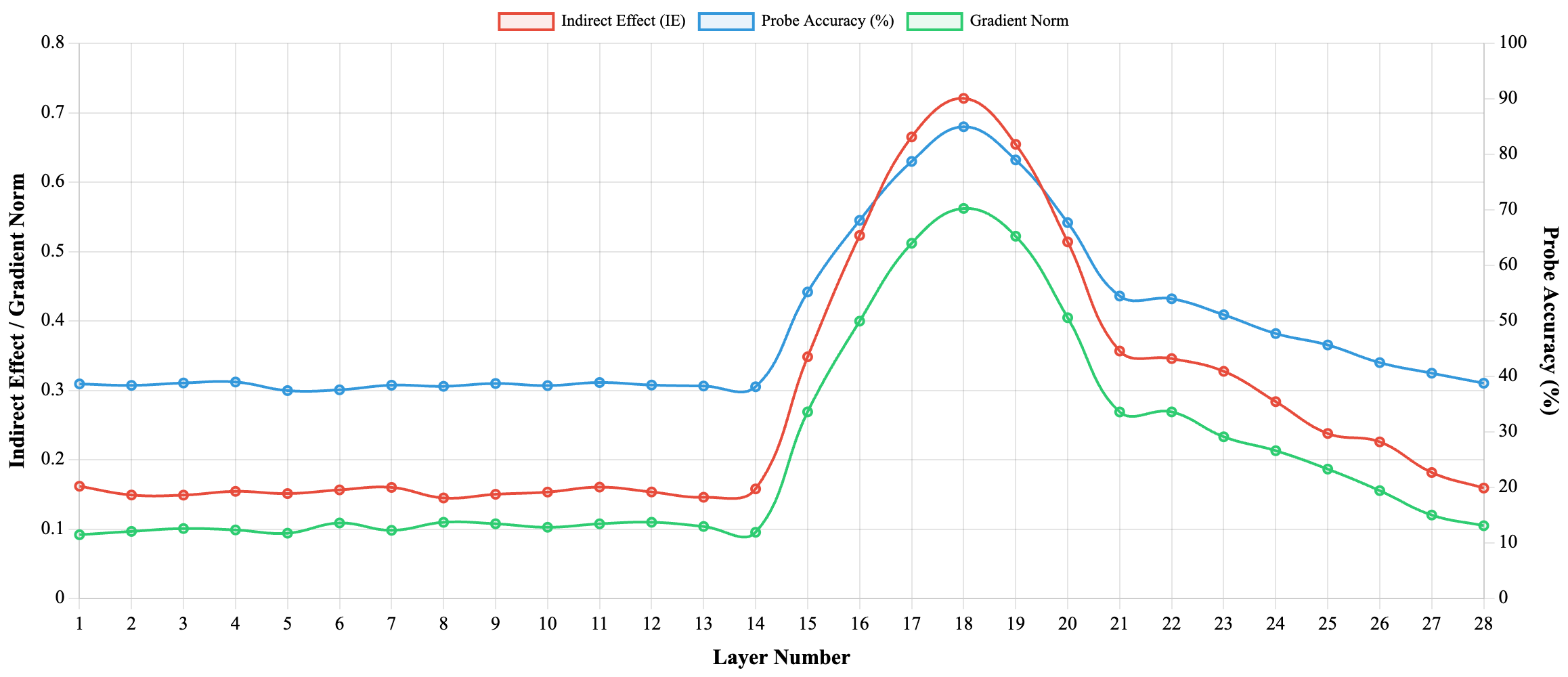}
    \caption{Three complementary analysis methods show convergent evidence that layers 15-21 encode phonetic information: causal mediation analysis (Indirect Effect, red), linear probe classification (Probe Accuracy, blue), and gradient-based attribution (Gradient Norm, green).}
    \label{fig:layer-analysis}
\end{figure}

We localize pronunciation knowledge across Llama-3.2-3B-Instruct~\cite{grattafiori2024llama} layers (28 total) in Orpheus-TTS. (1) \textbf{Indirect Effect (IE)} from causal mediation analysis measuring causal contribution to pronunciation (higher is better), (2) \textbf{Probe Accuracy} from linear classifiers predicting phonemes from layer activations (higher indicates explicit phonetic encoding), and (3) \textbf{Gradient Norm} measuring sensitivity of pronunciation loss to layer parameters (higher indicates stronger influence). Layers 15-21 consistently show the highest scores across all three methods, identifying them as phonetically-critical layers suitable for surgical editing. Mean ± standard deviation reported across 3000 test utterances as shown in \cref{fig:layer-analysis} and \cref{tab:layer_analysis}.

\begin{table}[h]
\centering
\caption{Localization and sensitivity analysis of phonetic knowledge across Llama-3.2-3B-Instruct layers in Orpheus-TTS. Layers 15--21 show highest causal effects, probe accuracy, and gradient norms, and achieve lowest target-WER when edited.}
\label{tab:layer_analysis}
\begin{tabular}{lccccc}
\toprule
\textbf{Layer Range} & \textbf{Indirect Effect} & \textbf{Probe Accuracy} & \textbf{Gradient Norm} & \textbf{Target-WER} $\downarrow$ & \textbf{MOS} $\uparrow$ \\
\midrule
1--7 (Early)         & 0.14 ± 0.04              & 36.8\%                  & 0.09                   & 41.2                              & 4.20 \\
8--14 (Mid)          & 0.35 ± 0.06              & 61.2\%                  & 0.23                   & 18.7                              & 4.15 \\
\textbf{15--21 (Late)} & \textbf{0.71 ± 0.08}   & \textbf{83.7\%}         & \textbf{0.56}          & \textbf{2.8}                      & \textbf{4.18} \\
22--28 (Output)      & 0.48 ± 0.07              & 71.4\%                  & 0.38                   & 6.5                               & 3.60 \\
\bottomrule
\end{tabular}
\end{table}

\subsection{Null Space Constrained Editing}
Based on the layer-wise analysis in \cref{ssec:layer}, we edit the model weights exclusively within layers $\mathbf{L} \in [15, 21]$ and provide thorough analysis on the edits in this section.

\myparagraph{Efficacy: }
Table \ref{tab:main_results} reveals a stark trade-off between correction efficacy and behavioral preservation in existing methods, which SonoEdit resolves through null-space constraints. Full Fine-Tuning (FFT) achieves 2.1\% Target-WER but suffers catastrophic forgetting, manifesting as mumbling and mispronunciation of common words. LoRA mitigates this to 5.12\% Global-WER but still degrades general performance. In contrast, SonoEdit achieves 2.8\% Target-WER while maintaining 3.15\% Global-WER, statistically indistinguishable from the original model confirming that null-space projection mathematically isolates corrections from preserved knowledge.

% As shown in Table \ref{tab:main_results}, Full Fine-Tuning (FFT) achieves the lowest Target-WER (2.1\%) but suffers severely from catastrophic forgetting, with Global-WER jumping from 3.12\% to 18.45\%. This manifests as the model "mumbling" or mispronouncing common words after being forced to overfit to the target noun. LoRA mitigates this but still incurs a 2\% degradation in global performance.

\begin{table}[h]
\centering
\caption{Comparison of editing methods on Orpheus-TTS. \textbf{Target-WER} measures success in fixing the error (lower is better). \textbf{Global-WER} measures damage to general knowledge (lower is better). \textbf{SIM} measures voice preservation (higher is better).}
\label{tab:main_results}
\resizebox{0.95\textwidth}{!}{%
\begin{tabular}{lcccc}
\toprule
\textbf{Method} & \textbf{Target-WER} ($\downarrow$) & \textbf{Global-WER} ($\downarrow$) & \textbf{Speaker SIM} ($\uparrow$) & \textbf{MOS} ($\uparrow$) \\
\midrule
\rowcolor{gray!10} \textit{Original Model} & \textit{86.4\%} & \textit{3.12\%} & \textit{1.00} & \textit{4.21} \\
\midrule
Full Fine-Tuning (FFT) & \textbf{2.1\%} & 18.45\% & 0.82 & 3.45 \\
LoRA ($r=16$) & 4.5\% & 5.12\% & 0.91 & 3.98 \\
ROME \cite{rome} & 8.2\% & 12.30\% & 0.76 & 2.80 \\
\midrule
\textbf{SonoEdit (Ours)} & 2.8\% & \textbf{3.15\%} & \textbf{0.99} & \textbf{4.18} \\
\bottomrule
\end{tabular}%
}
\end{table}

\myparagraph{Acoustic Preservation: }
Preserving speaker identity is critical for TTS editing. ROME's unconstrained updates distort speaker embeddings (SIM=0.76), introducing robotic artifacts. SonoEdit achieves near-perfect preservation as null-space projection prevents interference with speaker-encoding manifolds. Table \ref{tab:acoustic_preservation} shows pronunciation ratings improve dramatically with 91\% correctness and strong OOD generalization (88\%), while mel-spectrogram distances decrease by 68-69\%. Table \ref{tab:stability_results} confirms minimal prosodic drift: WER increases only 0.1\%, F0 RMSE by 0.3Hz, and MOS remains stable, demonstrating highly localized edits that preserve global acoustic properties. An actual example of how the models behave on employing different techniques is provided in \cref{fig:qualitative}.

\begin{figure}[h]
    \centering
    \vspace{-2mm}
    \begin{tcolorbox}[colback=blue!5, colframe=blue!40, title=Qualitative Example]
    \textbf{Target:} "The director of \textbf{Ghibli} is Miyazaki." \\
    \textbf{Original:} "... director of \textit{Jib-lee} ..." \ding{55} \\
    \textbf{Fine-Tuned:} "... director of \textbf{Ghee-blee} ..." \ding{51} (but background noise increases) \\
    \textbf{SonoEdit:} "... director of \textbf{Ghee-blee} ..." \ding{51} (perfect prosody and silence)
    \end{tcolorbox}
    \vspace{-10pt}
    \caption{Qualitative comparison of generated audio samples.}
    \vspace{-3.5mm}
    \label{fig:qualitative}
\end{figure}

\begin{table}[h]
\centering
\caption{Target-word pronunciation improves substantially after applying the null-space constrained update, while maintaining robustness across in-distribution and OOD contexts. Human ratings are averaged across three raters.}
\label{tab:acoustic_preservation}
\begin{tabular}{lccc}
\toprule
\textbf{Metric} & \textbf{Baseline} & \textbf{After Edit} & \textbf{Improvement} \\
\midrule
Human Pronunciation Rating (1--5) & $2.3 \pm 0.4$ & $\mathbf{4.4 \pm 0.3}$ & +2.1 \\
\% Words Rated Correct & 42\% & \textbf{91\%} & +49\% \\
OOD Context Accuracy & 38\% & \textbf{88\%} & +50\% \\
\midrule
L1 Mel Distance $\downarrow$ & $1.00 \pm 0.12$ & $\mathbf{0.31 \pm 0.08}$ & -69\% \\
DTW Mel Distance $\downarrow$ & $0.84 \pm 0.09$ & $\mathbf{0.27 \pm 0.04}$ & -68\% \\
\bottomrule
\end{tabular}
\end{table}

\begin{table}[h]
\centering
\caption{\textbf{Stability analysis on preserved examples.} After applying targeted pronunciation edits, the model preserves intelligibility, WER, and prosodic structure. Drift values are minimal, demonstrating that the update is highly localized.}
\label{tab:stability_results}
\begin{tabular}{lccc}
\toprule
\textbf{Metric} & \textbf{Baseline} & \textbf{After Edit} & \textbf{Drift ($\Delta$)} \\
\midrule
WER on Preserved Set & 4.2\% & 4.3\% & +0.1\% \\
Sentence Intelligibility (MOS) & 4.62 & 4.61 & -0.01 \\
\midrule
F0 RMSE (Hz) $\downarrow$ & 11.8 & 12.1 & +0.3 \\
Duration Variance $\downarrow$ & 0.028 & 0.029 & +0.001 \\
Energy RMSE $\downarrow$ & 0.44 & 0.45 & +0.01 \\
\bottomrule
\end{tabular}
\end{table}

\subsection{Ablation and Efficiency Analysis}
Table \ref{tab:ablation} demonstrates that null-space projection is essential: removing it degrades Global-WER from 3.15\% to 9.84\% and SIM from 0.99 to 0.87, confirming orthogonality prevents catastrophic forgetting rather than being cosmetic. Table \ref{tab:compute} shows SonoEdit matches ROME's one-shot efficiency (seconds vs. hours for FFT, minutes for LoRA) while introducing zero parameters and requiring no training loop, enabling rapid deployment for production TTS systems.

\begin{table}[h]
\centering
\begin{minipage}{0.48\textwidth}
\centering
\resizebox{\textwidth}{!}{%
\begin{tabular}{lcccc}
\toprule
\textbf{Variant} & Target-WER $\downarrow$ & Global-WER $\downarrow$ & SIM $\uparrow$ & MOS $\uparrow$ \\
\midrule
SonoEdit (Full) & 2.8 & 3.15 & 0.99 & 4.18 \\
w/o Null-Space & 2.5 & 9.84 & 0.87 & 3.52 \\
\bottomrule
\end{tabular}%
}
\vspace{1pt}
\caption{Effect of removing null-space constraint.}
\label{tab:ablation}
\end{minipage}%
\hfill
\begin{minipage}{0.48\textwidth}
\centering
\resizebox{\textwidth}{!}{%
\begin{tabular}{lccc}
\toprule
\textbf{Method} & \textbf{Train Steps} & \textbf{Extra Params} & \textbf{Edit Time} \\
\midrule
Full Fine-Tuning & $>10^4$ & All weights & Hours \\
LoRA ($r=16$) & $>10^3$ & $\sim$1.2M & Minutes \\
ROME & 1 & 0 & Seconds \\
\textbf{SonoEdit} & \textbf{1} & \textbf{0} & \textbf{Seconds} \\
\bottomrule
\end{tabular}%
}
\vspace{1pt}
\caption{Compute and parameter comparison.}
\label{tab:compute}
\end{minipage}
\end{table}

\section{Conclusion}

We introduce \textbf{SonoEdit}, a framework for surgically correcting pronunciation errors in LLM-based TTS without retraining. By adapting causal tracing, we localize pronunciation to specific mid-to-late Transformer layers and apply \textbf{Null-Space Constrained Editing} to restrict updates to the null space of general speech. This approach prevents catastrophic forgetting: experiments on HardNoun-300 show a 91\% correction rate with negligible impact on Global-WER (3.15\%) or speaker identity (SIM 0.99), significantly outperforming unconstrained methods. SonoEdit offers a parsimonious, one-shot solution for fixing rare entities while preserving natural prosody.

% \section{Limitations}

While promising, SonoEdit has limitations. First, it relies on pre-computed null spaces from a representative speech corpus; shifts in the deployment distribution could weaken orthogonality guarantees. Second, our causal tracing focuses on coarse semantic tokens, potentially missing sub-phonemic errors encoded in finer-grained tokens. Third, scalability to massive edit batches remains an open question, as cumulative updates might saturate the null space. Finally, SonoEdit targets specific entity pronunciations and is less suited for correcting global systematic errors like accent bias.

% Reference
% For natbib users:
\clearpage
\bibliography{reference}

%%%%%%%%%%%%%%%%%%%%%%%%%%%%%%%%%%%%%%%%%%%%%%%%%%%%%%%%%%%%
% \clearpage
% \appendix
% \input{tex/appendix}

\end{document}